\documentstyle[sao2,psfig]{article}
\renewcommand{\Delta}{\bigtriangleup}
\newcommand{\hal}{${\rm H}\alpha$}
\def\sec{\ifmmode {}^{\prime\prime}\else ${}^{\prime\prime}$\fi~}

\def\magdot{\ifmmode {}^{\rm m}\!\!\!.\, \else ${}^{\rm m}\!\!\!.\,$\fi}
\def\asec{\ifmmode ^{\prime\prime}\else$^{\prime\prime}$\fi}
\newcommand{\hbe}{${\rm H}\beta$}
\newcommand{\hga}{${\rm H}\gamma$}
\newcommand{\rou}{$\rm R_{out}$}
\newcommand{\rin}{$\rm R_{in}$}
\newcommand{\av}{$\rm A_v$}
\newcommand{\nh}{$\rm N_H$}
\newcommand{\wl}{${\rm W}_\lambda$}

\newcommand{\lbl}{${\rm L_{bol}}$}
\newcommand{\lk}{${\rm L_k}$}
\newcommand{\tetj}{${\rm \theta_j}$}

\newcommand{\vt}{${\rm V_t}$}
\newcommand{\vj}{${\rm V_j}$}
\newcommand{\vc}{${\rm V_c}$}
\newcommand{\lm}{$\lambda$}
\newcommand{\ic}{${\rm I_c}$}
\newcommand{\tbsp}{\rule{0pt}{18pt}}
\issue{1997, 43, 37-54}
\begin{document}
\title{Spectral study of MWC\,560. Parameters of the system, the hot
source and the jets}
\author{
A.A. Panferov\inst{a} \and S.N. Fabrika\inst{a} \and  T. Tomov\inst{b}}
\institute{\saoname \and
National Astronomical Observatory Rozhen,
P.O. Box 136, 4700 Smoljan, Bulgaria}
\date{March 21, 1997}{May 12, 1997}
\maketitle

\begin{abstract}

Results of spectroscopy of unusual symbiotic variable star MWC\,560 
are presented.
The data have been obtained from autumn 1990 to spring 1993. High
velocity absorption lines of H\,I, ${\rm D_1\,D_2}$ Na\,I, Fe\,II(42)
and He\,I are permanently presented in the spectra, their radial
velocity is $-500 \div -2500$~km/s and is changed with a time--scale
of a few months. The high velocity absorption line profiles demonstrate
their origin in a jet, directed closely to the line of sight. These
absorptions are variable at a time-scale less than one day and with
an amplitude of about 10~\%. The fast variability of the absorption line
profiles at $\la 3$~hours have been observed. An evidence of the opposite
(second) jet has been found --- there is an additional and variable
emission in a red wing of the \hal\ emission line profile. The radial
velocity of this emission corresponds to the jet velocity, measured
from the absorption lines.

An estimates of the orbital inclination angle (${\rm i \approx 10^{\circ}}$)
and the jets parameters have been derived. We have simulated the continuum
spectrum from UV to IR in the two different states (high and low)
of MWC\,560. The observed spectrum is formed in a M4--5III giant,
a hot source and in an absorbing screen (the jet). The interstellar
absorption is ${\rm A_V = 1\magdot4 \div 1\magdot6}$. The jet opening
angle 2\,\tetj \,$\ga 40^{\circ}$, the jet length in the place of the
absorption is $\approx 1\cdot10^{12}$~cm. The absorbing gas temperature
and density are
${\rm T_j \approx 8000}$~K and ${\rm n \approx 5\cdot10^{11}\,cm^{-3}}$.
The jet column density is variable and equal to
${\rm N\approx 3\div6 \cdot 10^{23}\,cm^{-2}}$. At different states of
the object only the jet velocity is changed (and, accordingly, N), but
the mass loss rate in the jet is about constant and equal to
${\rm \dot{M}_j\approx5\cdot10^{-7}\, M_{\odot}/y}$. The object
luminousity ${\rm L_{bol} \approx 2\cdot 10^{37}}$~erg/s is radiated mainly 
by the UV source, whose temperature and size are ${\rm T \approx 20000}$~K
and ${\rm R \approx 4\cdot10^{11}}$~cm. The hot source is probably a
photosphere of a massive wind (${\rm \dot{M}_j\approx2\cdot10^{-6}\,
M_{\odot}/y}$, ${\rm V = 100\div200}$~km/s) rising from inner area
aroung the white dwarf.

We suppose that both the wind and jets are accelerated in magnetic
field of the white dwarf, which accretes the gas from the giant wind
as a propeller. The gas flows in accretion disk and is expelled from that
by the rotating magnetic field. A sectorial structure of the outflow (jets)
is formed. The different states of MWC\,560 could be corresponding to
different propeller regimes: a hard propeller with a gas velocity outflow
a few thousands of km/s and a soft propeller with a few hundred km/s.
A total mass accretion rate onto the propeller is
${\rm \dot{M} \approx 3 \div 4\cdot10^{-6}\, M_{\odot}/y}$.
A main part of this mass flux (about 50 -- 70~\%) is outflowed as a
slow wind, $\la$ 15~\% could reach the white dwarf surface and about
20 -- 30~\% is expelled as a jets. A surface magnetic field strength on
the white dwarf is ${\rm 3\cdot10^{8}\,G \la B_S \la 3\cdot10^{9}\,G}$,
its rotational period is from 10 -- 20 minutes to 3 hours.
\end{abstract}

\section{Introduction}
MWC\,560 is a symbiotic binary star with an unusual spectrum
distinguished by strong variables absorption lines of hydrogen
and singly ionized metals. The absorption components of the
lines in the spectrum of MWC\,560 are detached (like BAL quasars)
in a sense that there is a portion of undistorted continuum between
the absorption and the emission lines. Tomov et al. (1990) have found
a shift of these absorptions up to $-6000$~km/s, which is
close to the parabolic velocity of the white dwarf, a likely
component in the system. Tomov et al. (1990) have proposed a
model of jet in MWC\,560 directed along the line of sight, in
which high--velocity absorptions are formed. For the detached
absorption components to appear, the radial velocity dispersion
in the jet must be considerably lower than the jet spread
velocity.

In symbiotic stars jets are known to be observed (Livio, 1996).
It is supposed that one of the principal conditions for generation
of jets is the presence of an accretion disk (Livio, 1996).
A photometric flickering (Tomov et al., 1996) --- a typical feature of
all cataclysmic variables in which gas accretion by a white dwarf
with a magnetic field occurs (Robinson, 1976) --- is evidence of the accretion
in MWC\,560. It is apparent that for the jet gas absorption lines
to appear in the spectrum, the gas must be projected onto a continuous
spectrum source, that is, the MWC\,560 accretion disk axis and therefore
the orbital axis of the binary system must be close to the line of
sight. The absence of noticeable variations in the radial velocities
of the emissions lines (Tomov et al., 1994) does suggest such an
orientation of the MWC\,560 orbit.
 A host of absorption lines of the jet (of Fe\,II and other
elements with a low excitation potential) form a pseudo-continuum
in a UV spectrum. From the modelling of this spectrum Shore et al. (1994)
have obtained the column density in the jet to be $10^{23}$~cm$^{-2}$
for this object being at a maximum UV luminosity (in the spring of 1990)
and $10^{24}$~cm$^{-2}$ at a minimum (in the autumn of 1990).

In this paper we present a spectral monitoring of MWC\,560
in 1990--1993 and describe the behaviour of the high-velocity
absorption lines, that may be essential for the understanding of
their origin. We show that the observed spectrum of the object is
formed with involvement of absorption in the jet both in the lines
and in the Balmer continuum. Further we model the continuous
spectrum with allowance made for the jet absorption and find the
parameters of the hot component and the jet.

\section{Spectral monitoring of
 MWC\,560. Variability of absorption lines}
From October 1990 to February 1993 we obtained about 50 spectra of
MWC\,560 on the 6-m telescope. The observation were carried out with the
spectrograph SP-124 and the IPCS --- a TV scanner (Drabek et al., 1986)
and with the echelle spectrograph ''Zebra'' and the panoramic IPCS ''Quantum''
(Gazhur et al., 1990). A spectral range of 3700--7000 \AA\AA\ was recorded
with a resolution \lm/$\Delta$\lm\ $=3500$ and 2300, an average exposure
time for a single spectrum of 20$^{\rm m}$, and a signal-to-noise ratio
of 10--30. The time distribution of the observations allowed us to study 
a spectral variability from 10$^{\rm m}$ to a week.

In accordance with Tomov et al. (1994) we use the numbering of observing
seasons which are determined by visibility of the object for ground-based
observations. The first season refers to the first half of 1990, when
the absorption lines in the spectrum of MWC\,560 showed a maximum
velocity and suffered a strong variability from night to night (even
disappeared) (Tomov et al., 1990). Our observations refer to seasons 2--4.
The dates of the observations and the mean parameters of the high-velocity
absorption line \hbe\ in these seasons --- the equivalent width \wl,
the residual intensity \ic, the terminal velocity \vt, determined from
the blue wing, the velocity \vc\ of centre of mass of the line, and
the width at half intensity FWHM --- are given in Table 1.

A few typical for these seasons spectra in the region containing \hga and
\hbe lines are shown in Fig.~1. The spectra reduced for a flat field are 
given in readout IPSC units. It is seen in the figure the high--velocity 
absorption lines behaviour a time goes on.

\begin{table*}
\begin{center}
\caption{Parameters of \hbe\ absorption in 1990--1993}
\begin{tabular}{c|c|c|c|c|c|c}
\hline   
  \multicolumn{1}{c}{}
                       & \multicolumn{1}{|c}{}
                       & \multicolumn{1}{|c}{}
                       & \multicolumn{1}{|c}{}
                       & \multicolumn{1}{|c}{}
                       & \multicolumn{1}{|c}{}
                       & \multicolumn{1}{|c}{}\\
\hline
season    & date of       & ${\rm I_c}$ & \wl\ & \vt\ & \vc\ & FWHM \\
$\cal N$ & observations   &      & \AA\ & km/s & km/s & km/s\\[.6ex]
\hline
2 & 9.10.90 -- 15.10.90 & 0.1 & $8\div 11$ & $900\div 1400$ & $400\div 500$ &
$500\div 700$ \\
  & 15.04.91            & 0.2 & 4.1        & 580            & 190           &
320           \\
\hline
3 & 29.10.91 -- 21.3.92 & 0.1 & 15         & $2500\div 2800$ & $1700\div
 1900$ &  1000 \\
\hline
4 & 21.09.92 -- 10.12.92 & 0.1 & $15\div 20$ & $2500\div 2900$ & $1600\div
 1900$ & $1000\div 1300$ \\
  & 9 -- 10.02.93 & 0.1 & $9\div 12$ & $1700\div 1900$ & 1200 &
 $600\div 800$\\
\hline
\end{tabular}
\end{center}
\end{table*}
\begin{figure*}
\centerline{\psfig{figure=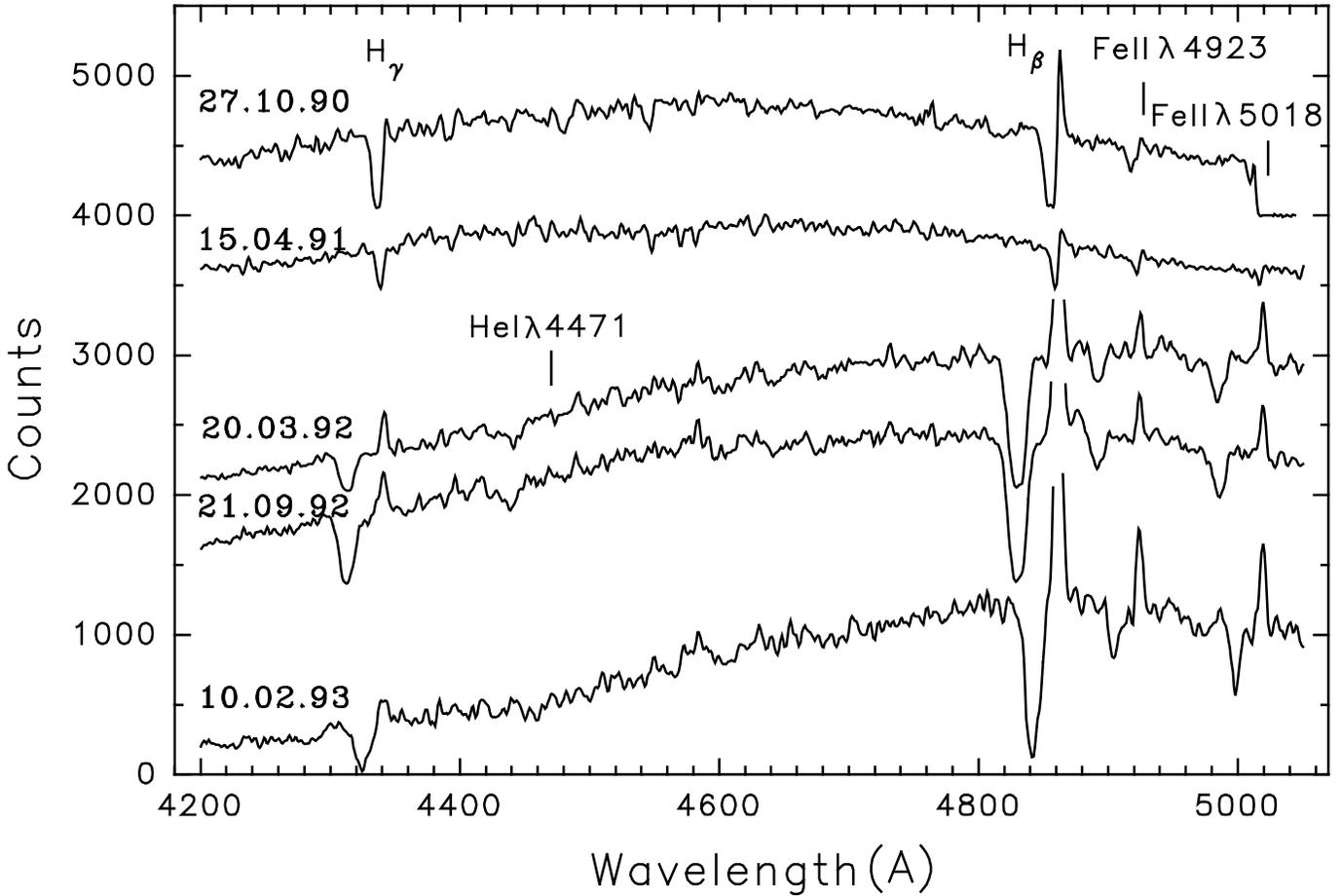}}
\caption{
Spectra of MWC~560 within 4200--5040~\AA, obtained at different seasons. The
dates of observations are shown. Counts are indicated on the vertical
axis. From top to bottom the spectra are shifted in intensity 
by 4000, 3400,
2000 and 1300 units. Emission of \hbe line is cut on three lower spectra}
\end{figure*}

Strong emission lines of H\,I and Fe\,II(42), and
deep variable absorption lines of these elements shifted bluward to
$- 3000$~km/s are prominent in the spectrum of MWC\,560. Besides those,
high--velocity absorptions of D$_1$ and D$_2$ Na\,I, He\,I and H and K
Ca\,II lines are also presented in our spectra, the latter being as deep
as the absorptions of H\,I lines (see also Tomov et al., 1992). In the
3d and 4th seasons broad and very deep H\,I absorptions lines are observed.
In the 2nd and at the end of the 4th seasons the absorption line profiles
resemble the P\,Cyg--type profiles, which appeared as a result of decreasing
gas velocity in the jet. The equivalent width of absorption lines
\hal, \hbe\ and \hga\  are approximately equal, which suggest the absorption
lines to be saturated, whereas the equivalent width of the emissions
of these lines are an order different from one another. The widths
of the emissions are 4--5 times less than those of the absorptions. 
The observed profiles, for instance, of the \hga\ line are unlike 
those calculated
for a spherically symmetric outflow (e.g. see Castor and Lamers, 1979).
The equivalent width ratio of the \hga\ absorption and emission lines is
smaller than 0.1, which requires the optical thicknes $\tau < 1$ for a
spherical wind (Castor and Lamers, 1979), nevertheless from the great
absorption line depth it can be concluded that $\tau > 1$. The observed
absorption lines do not resemble the absorptions formed in an anisotropic
wind either as, for example, in the cataclysmic variable RW\,Sex, in which
the wind velocity amounts  to 4300~km/s (Greenstein and Oke, 1982).
So, the high--velocity absorption lines in the spectrum of MWC\,560 may be
formed in a jet directed to the observer.

In Fig.~2 are shown the absorption line profiles of different elements,
which have been obtained from the echelle spectrum of December 9, 1992.
It is seen that the widths of the Fe\,II absorption lines are smaller
than those of the H\,I absorption lines. The absorption lines 
Fe\,II~\lm\,4924 and \lm\,5018 may be blended with the absorption lines 
He\,I~\lm\,4922 and \lm\,5015, respectively. The latter, however, are
much weaker than the lines of iron since the intensity of other
He\,I absorption lines are small. It can be seen from Fig.~2 that the
jet is inhomogeneous in both velocity and temperature.

\begin{figure*}
\centerline{\psfig{figure=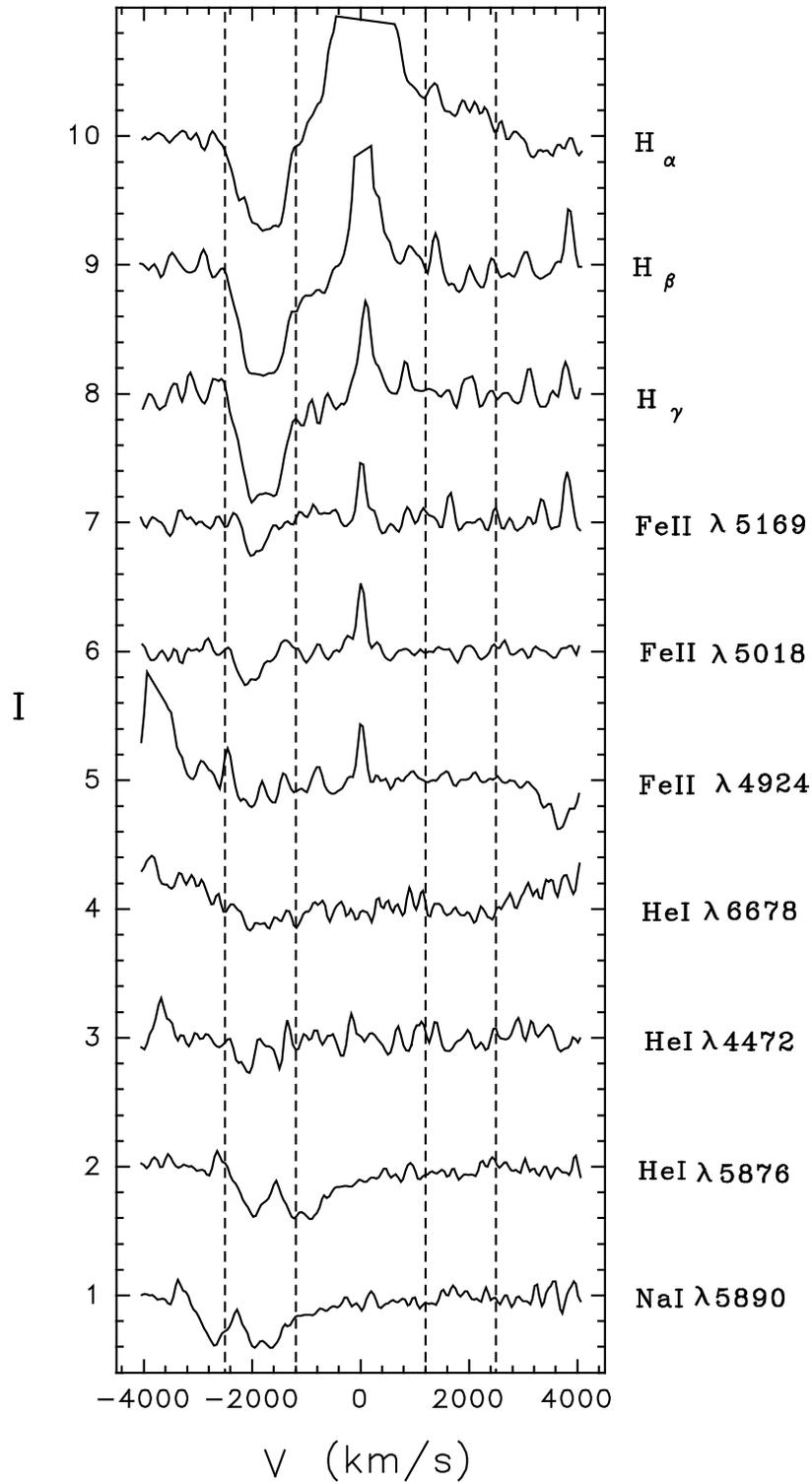,height=20cm}}
\caption{Profiles of high--velocity absorptions, obtained on December 9, 1992
(UT = 9.931). Horizontal axis shows the velocities with respect to
the laboratory wavelength, the vertical -- residual intensities.
The dashed lines are given for a convenience of comparing these
profiles. \hbe\ and \hal\ emissions are cut.}
\end{figure*}

The intensities \ic\ of the absorption lines H\,I are the same for
different seasons, but for the end of the 2nd season. The decrease
in \ic\ and FWHM of the absorption lines in the spring of 1991 (Table~1)
is due to the fact that the absorption lines are approached the
emission lines. High excitation lines typical of symbiotic variables have not
been revealed in our spectra.

The time resolution for the study of variability from our
spectra on different dates was from 10$^{\rm m}$ to 5$^{\rm h}$.
The quality of spectra on some dates was sufficient for 3~\% of variability
to detect. A daily variability of high-velocity absorptions with an
amplitude of 10--20~\% was always observed, which is seek from Fig.~3,
where  the behaviour of the \hga, \hbe\ and \hal\ absorption lines
from the spectra taken from December 4 to 10, 1992 is presented.

\begin{figure*}
\centerline{\psfig{figure=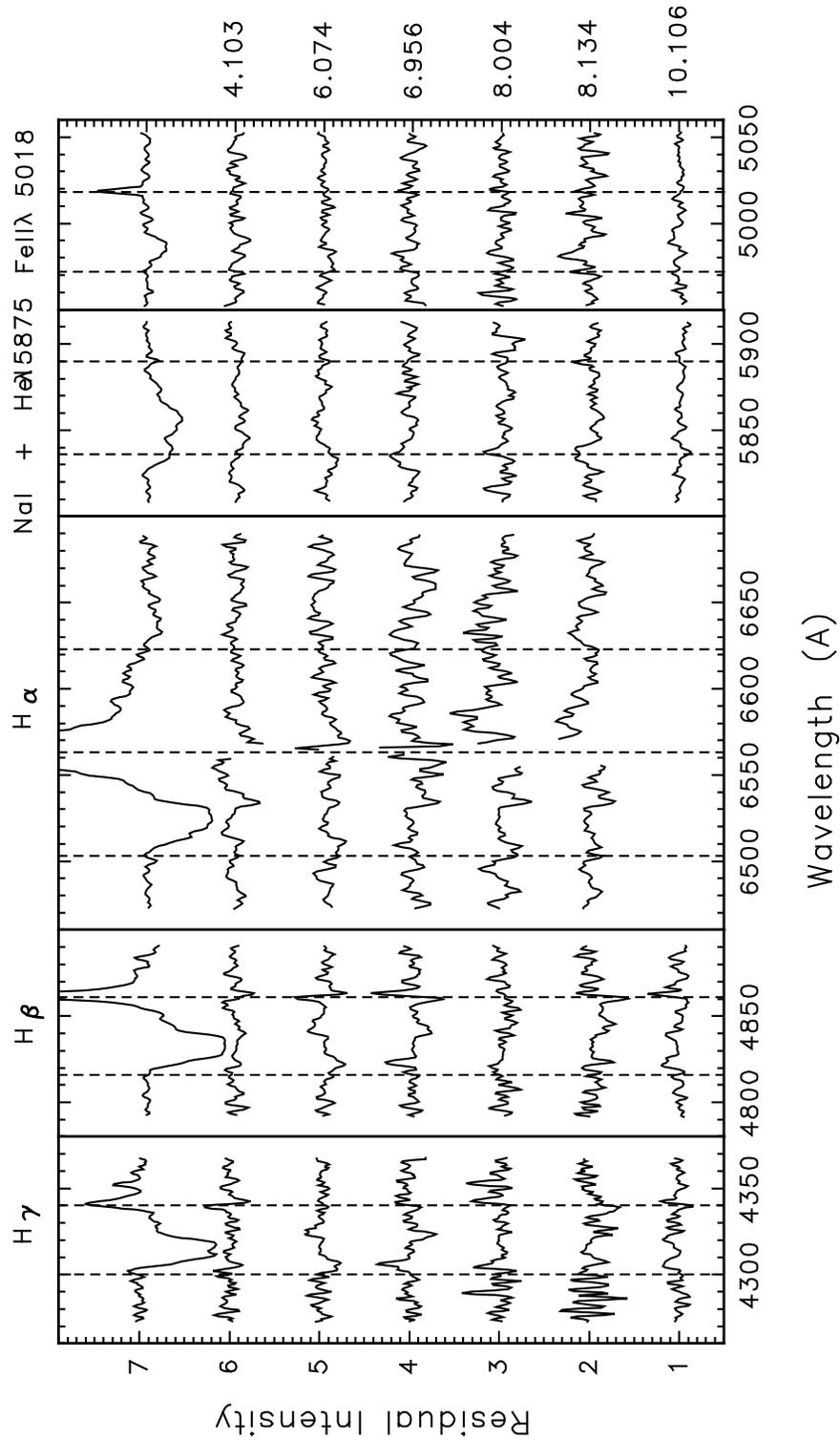,angle=90,height=20cm}}
\caption{
The summary line profiles  \hga, \hbe, \hal, \,${\rm D_1,D_2
\,Na\,I+He\,I \,\lambda\,5875, \,Fe\,II\, \lambda\,5018}$ 
and their differences with those of December 4--10, 1992. All the spectra
are normalized. A daily variability is noticed. The spectra on December 
8 show absorption variability for 3 hours. The time of observations (UT) 
is indicated right. The vertical dashed lines are given for a convenience 
of the profiles comparing. }
\end{figure*}

To demonstrate the variability, we show the differences between
the spectra under the study and a reference spectrum, which is a
sum spectrum over all the observations (at the top in Fig.~3).
However the fragment of the reference spectrum with the \hal\ line is presented
from the spectrum on December 9 to show more vividly the broad red wing
of this line (manifestation of the counter jet, see below). All the
spectra have been reduced to residual intensities. The vertical dashed
lines indicate a zero radial velocity and $-2800$~km/s ($\pm 2800$ for
the \hal). The profile variability of the high-velocity absorptions can
be seen in all the spectra in the lines \hal, \hbe\ and \hga. The
variability of the absorptions in the spectra on December 8, which
were obtained with an interval of 3 hours, can also be seen in the
figure: the change in the intensity occurred simultaneously for the
absorption lines \hga\ and \hbe\ at ${\rm V_r=-2360}$ and --1090~km/s.
At the same time no changes in \hal\ absorption were noticed,
possibly because of the great optical thickness of this line. In the
examples of the night-to-night variability, \hal\ is markedly different
from the lines \hga\ and \hbe, whose absorption line profile variability is
practically synchronous. The amplitude of the 3-hour's variability
in \hga\ and \hbe\ is not smaller than that of the night-to-night
variability. In spite of the fact that occasionally the variability
amplitude slightly exceeds the noise in the spectra the apparent
synchronism of the variations in the \hga\ and \hbe\ absorption line
profiles makes this variability significant. On other nights no
fast variability was noticed, that is why the differences for 
night's sum spectra are only presented for these nights. A variability
of high--velocity absorptions in the UV spectra at times of $\le 3$
hours has been detected by Michalitsianos et al. (1991), while at times
of about 1 hour --- by Fabrika et al. (1991). Based on this variability we can
estimate the jet length. Taking the characteristic time of the variability
to be ${\rm \Delta t \la 3}$ hours and the gas velocity in the jet during the
observations on December 1992 ${\rm V_j\approx 2000}$~km/s we find
${\rm R_j \le \Delta t V_j \approx 2\cdot 10^{12}}$~cm.

A broad red wing in the \hal\ emission line is well visible on the spectra;
its width is comparable with the width of the absorption line (Fig.~2).
The wing also shows variability from night to night (Fig.~3). We
suppose that in the red wing of the \hal\ line the radiation of the jet 
directed away from us (the counter jet) manifests itself. 
This radiation may be shielded
partially by the hot component of the system (accretion disk). The excess
in the red wing above the \hal\ line core profile approximated by
Gaussian has an equivalent width of about 4 \AA. The total luminosity
in this wing is L\,(\hal) $\ga 2\cdot 10^{33}$ erg/s. Here we have
adopted the mean values of the following parameters (see Table~2 and
Fig.~5 below): the distance to MWC\,560 --- 1.7 kpc, the interstellar
extinction and the flux in continuous spectrum near \hal\ --- about 
1${\rm ^m}$ and ${\rm F_\lambda = 5\cdot 10^{-13}}$~erg/cm$^2$\,s\,\AA. 
On the basis of
the specific emission measure in the \hal\ line ${\rm \approx 2\cdot 10^{-26}
N_e^2}$~erg/s\,cm$^3$\,ster, for the expected temperature in the jets
$\approx 10^4$~K (see Table~2), we obtain the emission measure:
${\rm V\,n_e^2 \ga 6\cdot 10^{57}}$~cm$^3$, where V is volume, 
${\rm n_e}$ is the electron density. Here the inequality conforms to the fact
that a part of the emission region is screened by the accretion disk and also
that the optical thickness of the emitting gas may be $>1$. Assuming
that the jet is conical (${\rm V \approx \sin^2\theta_j\,R_j^3}$) and
using the jet length obtained from the time of
variability, we can estimate the column electron density in the jet
${\rm N_e = n_e\,R_j}$. Assuming that the semi--opening of the jet
in MWC\,560 is ${\rm \theta_j =20^\circ}$, we obtain 
${\rm N_e \la 1.6\cdot 10^{23}}$~cm$^2$. Below based on the modelling 
of the continuous spectrum of MWC\,560
we determine the total column density (the number of protons $+$ the
number of hydrogen atoms) of the absorbing gas in the jet, ${\rm N_H
\approx 10^{24}}$~cm$^{-2}$. This has been derived for the jet directed
toward us while the electron density ${\rm N_e}$ has been found from
observations of the counter jet. We have not any grounds to believe
that the two jets differ in their properties, therefore it follows
from these two values that the hydrogen ionization degree in the MWC\,560 jets
${\rm n_e/n_H \ga 0.15}$, i.~e. the gas temperature ${\rm T_j \approx 
7000\div 10000}$~K. Note that this estimate of the ionization degree  
in the jets does not depend on the volume filling factor of the gas, i.~e. on
whether the gas is in clouds or the whole volume of the jet
is completely filled with the gas.

The results of observations are in very well agreement with the idea that 
the MWC\,560 
jets oriented close to the line of sight. It is natural to suppose
that the jets are perpendicular to the accretion disk, i.~e. the axis of
the jets is parallel to the angular momentum of accreted matter.
From this it also follows that the orbital inclination angle is small.
From the data of Tomov and Kolev (1997) that the radial velocity of
the Fe\,II, Ti\,II and of other metal emission lines is about 35~km/s and
does not change in time; within an interval of more than 1000 days the
radial velocities change by less than 4~km/s. On the base of CCD
spectra taken later on the 6-m telescope E.\,L.~Chentsov (private communication,
1997) has reported the amplitude of radial velocity variations of the
emission lines to change less by 2~km/s, here again for a time of
more than 1000 days. The metal emission lines are produced as a result of
fluorescence of UV radiation (Shore et al., 1994) and, apparently,
belong to gas outflowing from the inner regions. Thus, we can
estimate the upper limit to the amplitude of orbital variability
of radial velocities as 2\,K $\la 4$~km/s. The orbital period of MWC\,560
determined by Doroshenko et al. (1993) from the data of archives and
up-to-date photometry is 1930 days. Assuming the masses of the white dwarf
and the giant to be ${\rm 1M_\odot}$ (see the discussion below), 
we find the orbital
velocity ${\rm V_{orb} \approx 10}$~km/s and the restriction on the possible
interval of the system's inclination angle value ${\rm \Delta i \le K/
V_{orb}\,\cos i \approx 10^\circ}$. The probability of finding a binary system
with i\,$=0^\circ$ is extremely low, therefore we may take i\,$ \approx
10^\circ$ for the inclination angle of the system.

From modelling the continuous spectrum of MWC\,560 (Table~2) we estimate
the size of the UV radiation source of about (4--5)\,$\cdot 10^{11}$~cm.
For the observed absorption lines in the jet to appear,
the size of the jet projection onto the sky plane (in the jet where 
the absorbent is situated, ${\rm R_j \la 2\cdot 10^{12}}$~cm) 
must be larger than that of the source. From this condition it
follows that ${\rm \theta_j - i > 10^\circ}$. Taking into account the 
inclination angle of the system (and the jet), i~$\approx 10^{\circ}$, 
we find that the semi--opening of the
jet is ${\rm \theta_j \ga 20^\circ}$. In the known symbiotic binary CH\,Cyg
this parameter determined from radio observations (Taylor et al., 1986)
is ${\rm \theta_j \approx 5^\circ}$. 

The MWC\,560 jet opening angle can also be
estimated based upon the jet absorption lines widths. This will be an upper
limit to the value since along with the dispersion of trajectories
of matter in the jet there are other mechanisms (e.~g. the dispersion of
gas ejection velocities) that may contribute to the absorption lines 
broadening.
The jet is inhomogeneous in both temperature and density. This follows
from the the variability of the high--velocity absorption lines and from
the fact that we observe absorptions of elements of essentially different
excitation potentials (Fe\,II, He\,I). Collision of individual fragments
of the jet at velocities of several hundred km/s must cause considerable
temperature variations of the jet gas. From examples of detached 
P\,Cyg profiles on the spectra of different dates in early 1990
(Tomov and Kolev, 1997) we find that the ratio of the minimal width
of the observed absorption lines to the maximal velocity
of ejection on the same dates is equal FWZM/\vt\ $\approx 0.2$. So we have
from this ${\rm \theta_j -i < 35^\circ}$ or ${\rm \theta_j < 45^\circ}$. 
Here the inequality
conforms to the fact that the absorption line broadening is caused not only
by the dispersion of trajectories of the gas motion but also by the
dispersion of gas velocities in the jet.

\section{Dependence of MWC\,560 brightness on the jet velocity}
The spectrum of MWC\,560 is formed in the M4--5 giant star 
and the hot source of spectral class B5--A0 (Zhekov et al., 1996). IR
radiation has been noticed to be slightly variable on time scale of
a few days, which is typical for late giants, although the mean brightness
has not changed for several years (Zhekov et al., 1996). At the same time
the optical brightness, especially the UV, is a subject to both rapid
variability (flickering) and more considerable long-time variations
(Tomov et al., 1996). Zhekov et al. (1996) have found that the colours
of the hot component do not agree with those expected from the
accretion disk. However they disregarded a possible absorption by matter
of the jet. At the column density of hydrogen in the jet ${\rm N_H =
10^{23\div 24}}$~cm$^{-2}$ (Shore et al., 1994) this absorption must be
essential in formation of the observed spectrum in the UV and even in
optics. Therefore, the variable mass flux in the jet (or even the
variable gas velocity in the jet at a constant mass flux) may be the
cause of the radiation flux variability.

In Fig.~4 are shown the time dependences of the velocity \vc\ and radiation
fluxes of the object in the V band (Tomov et al., 1996) and in the UV.
A correlation is well seen in the velocity variations of the gas ejection
and the UV flux. The flux from the object in optics also shows a similar
variations, however the amplitude of the flux variations in the V band is small,
and the relation between the two quantities may be more complicated.

\begin{figure*}
\centerline{\psfig{figure=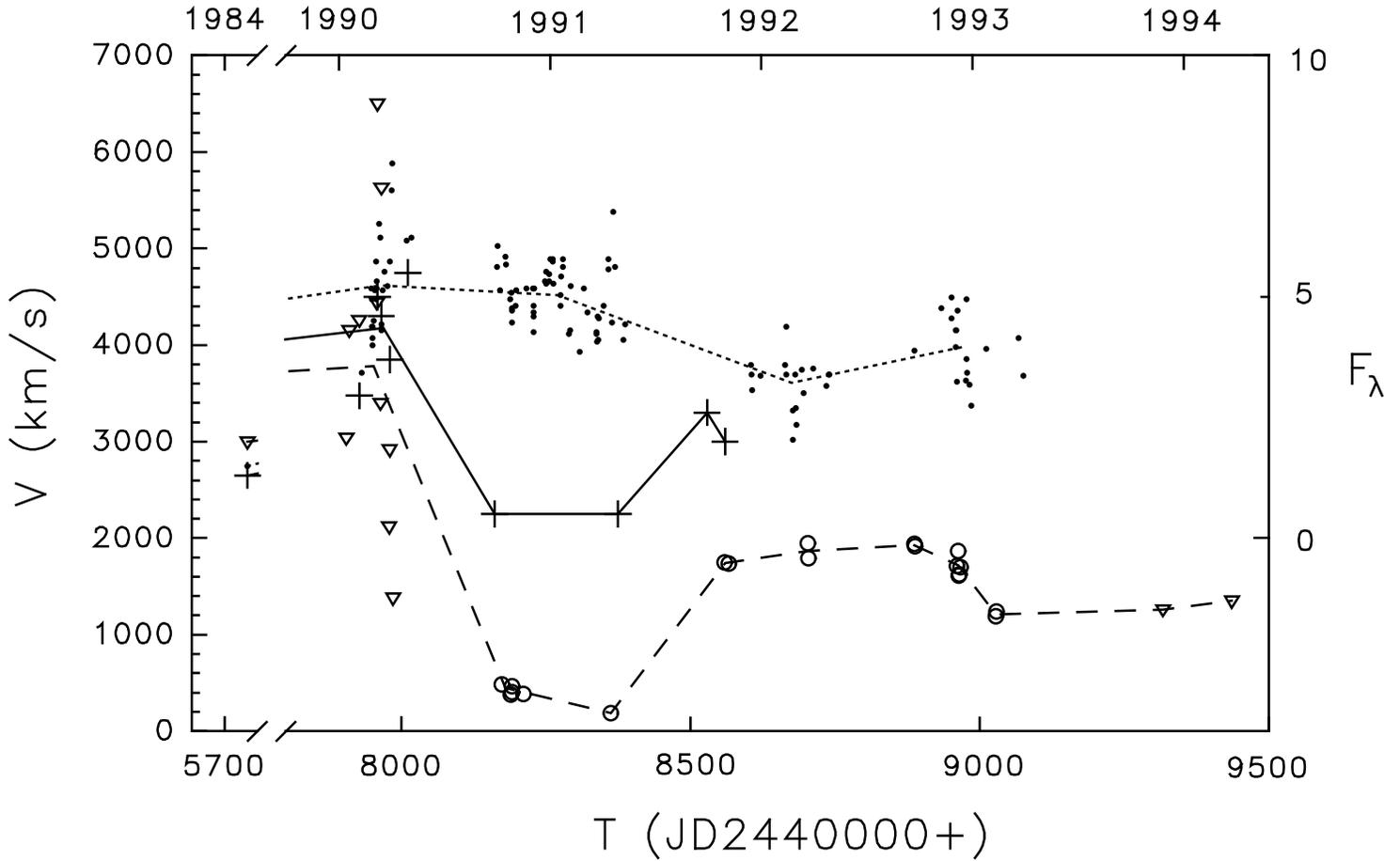}}
\caption{Time variations of MWC~560 activity. The \hbe\ line center of
mass velocity ${\rm V_c}$ is shown (circles --- our data, triangles --- 
Tomov et al. (1990; 1992; 1994)). Radiation fluxes in optics --- dots, 
in UV --- crosses. The axes show: left --- the velocity in km/s, right --- 
the radiation flux in ${\rm 10^{-13}~erg\,
cm^{-2}\,s^{-1}}$~\AA. Bottom --- the time in days (JD\,2440000+), top --- in
years.}
\end{figure*}

By the fall of 1990, in the 2nd observing season the mean outflow velocity
in the jet dropped to \vc\ $\approx 500$~km/s and became even lower in the
spring of 1991 (\vc\ $\approx 200$~km/s). This was accompanied by a
decrease in brightness to $0\magdot5$ and by the reddening 
${\rm \Delta(U-B) \approx 0\magdot5}$
and a more dramatic fall in UV radiation intensity (compared to the
spring of 1990). In the 3d season by the autumn of 1991 the velocity
increased and kept increasing till 1992. In 1992 the luminosity was
growing as well. Between 1992 and 1993 the velocity was observed to
decrease a little at the constant brightness of the star. Further
the velocity did not change much. In the period under study the object
MWC\,560 was in a passive state, when the absorption lines had a minor
variability as distinct from the very active state in the spring of 1990,
when the absorption lines were observed to be strongly variable (Tomov
et al., 1990). The brightness variations in the passive state are also
smaller than in the active one.

In the 2nd season in the spring of 1991 the velocity of the jet was
by approximately a factor of 5 lower than the mean velocity in the periods
that followed and by an order lower than the velocity in the 1st season.
The cardinal changes in the UV spectrum took place too (Fig.~4). The total
UV intensity in the 2nd season was lower, than both in the 1st and in
subsequent seasons, about 5--6 times as small, while in the range shorter
than 1400 \AA\ the difference was still greater (Maran et al., 1991)
and amounted to several dozen times. The UV spectrum in September 1990,
when the velocity of gas in the jet dropped to $\approx 1000$~km/s, was
similar to those of classical novae several days after the maxima.
This type of spectra has been called ''the iron curtain'' (Shore et al.,
1994): powerful unidentifiable emission features, which, in fact, are the
portions of the continuum that remained after absorption of the  strong
UV multiplets of Fe\,II and other ions of low ionization potentials
in moving matter. Such dramatic changes in the UV were practicaly
not reflected in the optical range, where these ions do not absorb. The strong
variations in the UV were thus caused by the variations of the optical
thickness of the absorbing gas, the real UV source luminosity was likely to be
unchanged. Since the redistribution of energy has not been observed,
then this gas is optically thick only in the directions close to the line
of sight, which is consistent with the jet outflow model. The variation of
column density of absorbing gas between different states of UV brightness
(Shore et al., 1994) reflects the gas mass variation on the line of sight
within the zone of ionization of principal absorbing agents. We  suppose
that the decrease in velocity of gas observed in the 2nd season caused
the increase in ${\rm N_H}$. Thus, physical conditions in the jet affect 
the shape
of the UV spectrum of MWC\,560. This will be shown by the spectrum modelling.

\section{Modelling the continuous spectrum of MWC\,560}
From the absorption lines of Ti\,O Thakar and Wing (1992) have defined the
spectral type of the giant in MWC\,560 as M\,4--5. When modelling, we have
used colours and absolute magnitudes of giants M4 and M5 from the data of
Lang (1992). Optical and UV spectra of symbiotic stars are peculiar,
and for their explanation 4 models a hot radiation source are usualy used:
1) a hot star in a planetary nebula, 2) a hot surface of a white dwarf,
at which permanent nuclear burning of accreted matter occurs, 3) an
accretion disk, 4) a hot optically thick wind from a white dwarf.
The models 1 and 2 are inconsistent with the fact that high excitation
lines are not observed in the spectrum and with the high luminosity
of the object (Table~2). Models 3 and 4 are likely to be more suitable
for the discription of MWC\,560. The observed flickering may be evidence
of a gas accretion onto a star with a magnetic field (Robinson, 1976). 
The accretor in MWC\,560 is, apparently, a white dwarf since the 
maximal observed velocity of the gas outflow (6000~km/s) close the 
parabolic velocity of a white dwarf.

The observed luminosity of MWC\,560 is an order lower than critical 
Eddington luminosity. 
This suggests that the supercritical accretion disk can not be considered
as the source of the jets. In the case of model 4 the wind may be due to
thermonuclear bursts of matter accumulated at the white dwarf surface
in the process of accretion or a rotating magnetic field of the white
dwarf acting like a propeller (Illarionov and Sunyaev, 1975). We have
considered the models 3 and 4 for the hot component as an accretion disk in
geometrically thin and locally blackbody approximations (Shakura and
Sunyaev, 1973) with a boundary layer (Lynden-Bell and Pringle, 1974)
and as a sphere with blackbody radiation. The jet been modelled as a
homogeneous hydrogen layer in a LTE state, which absorbs radiation from
the hot source. The velocity of gas motion was assumed equal to 1000~km/s. 
This velocity does not affect the shape of the continuum, but only
slightly changes the position of the Balmer jump. For the assumed
temperatures of gas in the jet of MWC\,560 (5000--10000~K, see above and
also Shore et al., 1994) the following basic processes of absorption and
scatter of continuous radiation have been taken into account: free-bound and
free-free transitions of atoms and negative hydrogen ions, Rayleigh
scattering by hydrogen atoms and Thompson scattering by free electrons.
The model spectrum of continuous radiation has been obtained with
allowance for interstellar extinction for which the approximations of
Cardelli et al. (1989), O'Donnell (1994) and the value ${\rm R_V=A_V/E_{(B-V)}
=3.14}$ have been used.

The model spectrum was compared with the observed spectrum which has
been obtained from the UBV data of Tomov et al. (1996), JHK data of Zhekov
et al. (1996) (see references therein) and UV data of IUE observations
of Michalitsianos et al. (1991) and Fabrika et al. (1991). From the
UV data the fluxes near \lm\,1470 and 3100 \AA\ have been only used, where
the continuum radiation is clearly seen between the strong absorption bands.
We have chosen two UV flux states of MWC\,560 (and corresponding to them 
the optical fluxes, see Fig.~4): with the maximal flux ---
in the spring of 1990 (Michalitsianos et al., 1991) and with the flux
typical for the main, more quiet, state of the object (Fabrika et al.,
1991) --- in the autumn of 1991. These two states correspond to the active
and passive periods (Table~1) determined above on the basis of gas
outflow velocities and brightness of the object. They are also different
in variability of the star: in the active state the object is more variable.
From examination Fig.~4 and from analyzing of the moving absorption line
profiles in different seasons (e.g. Tomov and Kolev, 1997) it can
be concluded that MWC\,560 actually exhibits a whole set of states. 
The transitions from one state to another may be rapid (it can be 
only said about the transition
time that it is shorter than a few months), but the characteristics of the
object (velocity in the jets, UV and optical brightness and variability 
amplitude)
change from a state to state continuously. So it could be more correctly to
say about the object activity levels in different observational seasons
than about the active or the passive states. Nevertheless, for modelling
the spectrum we have chosen two seasons corresponding to the two
states of the object, hereafter passive and active.

\begin{table*}
\begin{center}
\caption{Models}
\begin{tabular}{p{4cm}|r|r|r|r|r|r|r|r}
\hline  \multicolumn{1}{c|}{Parameters \tbsp}
                       & \multicolumn{2}{c}{ad1}
                       & \multicolumn{2}{|c|}{bb1}
                       & \multicolumn{2}{c}{ad2}
                       & \multicolumn{2}{|c}{bb2} \\
\cline{2-9}
& M4\,III& M5\,III& M4\,III& M5\,III& M4\,III& M5\,III & M4\,III & M5\,III \\
\hline \tbsp
\par
D [kpc]                          & 1.43  & 2.00  & 1.43  & 2.00            & 1.42  & 2.00  & 1.42  & 2.00  \\
${\rm A_V}$(G)                         & 1.64  & 1.10  & 1.62  & 1.10            & 1.61  & 1.03  & 1.57  & 1.10   \\
${\rm A_V}$(H)                         & 1.26  & 1.17  & 1.46  & 1.44            & 1.27  & 1.12  & 1.29  & 1.57   \\
${\rm R_{in}}$ [$10^{10}$~cm]          & 4.4   & 7.3   & ---   & ---             & 4.3   & 7.8   & ---   & ---    \\
${\rm R_{out}}$ [$10^{10}$~cm]         & 48.7  & 67.7  & 38.6  & 54.1            & 41.5  & 58.7  & 36.3  & 54.8   \\
T [$10^4$~K]                     & 2.72  & 2.40  & 2.11  & 2.08
& 2.50  & 2.10  & 1.79  & 1.99   \\ $\dot {M}_a$ [$10^{-5}\ M_\odot/{\rm y}$]
                                 & 9     & 25    & ---   & ---             & 6     & 18    & ---   & ---    \\
${\rm L_{bol}}$ [$10^{37}$~erg/s]       & 1.73  & 2.90  & 2.10  & 3.90            & 1.18  & 1.95  & 0.96  & 3.35   \\
${\rm T_j}$ [$10^4$~K]                 & 0.78  & 0.78  & 0.78  & 0.78            & 0.68   & 0.68   & 0.78  & 0.78   \\
${\rm N_ H}$ [$10^{23}$~cm$^{-2}$]& 2     & 2.5   & 3     & 3               & 10    & 12    & 6     & 6      \\
n [$10^{11}$~cm$^{-3}$]          & 6     & 5     & 5     & 5               & 6     & 6     & 5     & 5      \\
$\chi^2$                         & 0.39  & 0.39  & 0.34  & 0.38            & 0.78  & 0.49  & 0.47  & 1.03   \\
\hline
\end{tabular}
\end{center}
\end{table*}

 In Table~2 are listed the parameters of the system derived from the models.
Designations of the models: ad --- an accretion disk, bb --- a sphere with
blackbody radiation, 1 --- active state, 2 --- passive state, M4\,III and
M5\,III --- spectral class of the giant star in the given model. Parameters
of the models: D --- the distance to the system, \av(G) and \av(H) ---
interstellar extinction in the V band determined from the spectra of the
giant and the hot component, respectively, ${\rm R_{in}}$ --- the inner radius of
the accretion disk, ${\rm R_{out}}$ --- the outer radius of the disk or 
radius of
the sphere, T --- the temperature of accretion disk on radius of 2.25\,${\rm
R_{in}}$ (where maximum contribution to the total luminosity is (Shakura and
Sunyaev, 1973)) or the temperature of the sphere, ${\rm \dot M_a}$ --- 
gas accretion rate, ${\rm L_{bol}}$ --- the bolometric luminosity of 
the hot component. ${\rm
T_j}$, ${\rm N_H}$, n --- the temperature, column and volume densities 
(hydrogen and protons) of absorbing gas in the jet. For the models with 
an accretion disk we have used 1\,M$_\odot$ as the white dwarf mass value, 
which is the
average white dwarf' mass in cataclysmic variables (Ritter, 1991).

In Fig.~5 are shown the data of observations and the simulated spectra 
in the models 
bb1 + M5\,III (solid line) and bb2 + M5\,III (dashed line) for the active and
passive states, respectively.  The vertical bars show uncertainties in the 
observational data.  The large scatter in the radiation fluxes in the optical
and UV ranges of the spectra in the active state corresponds to the actually
observed variability of the object.  The models are consistent with the
observations. The other models from Table~2 fit the observational
data as well. The discrepancies of the models $\chi^2$ are given in Table~2.

\begin{figure*}
\centerline{\psfig{figure=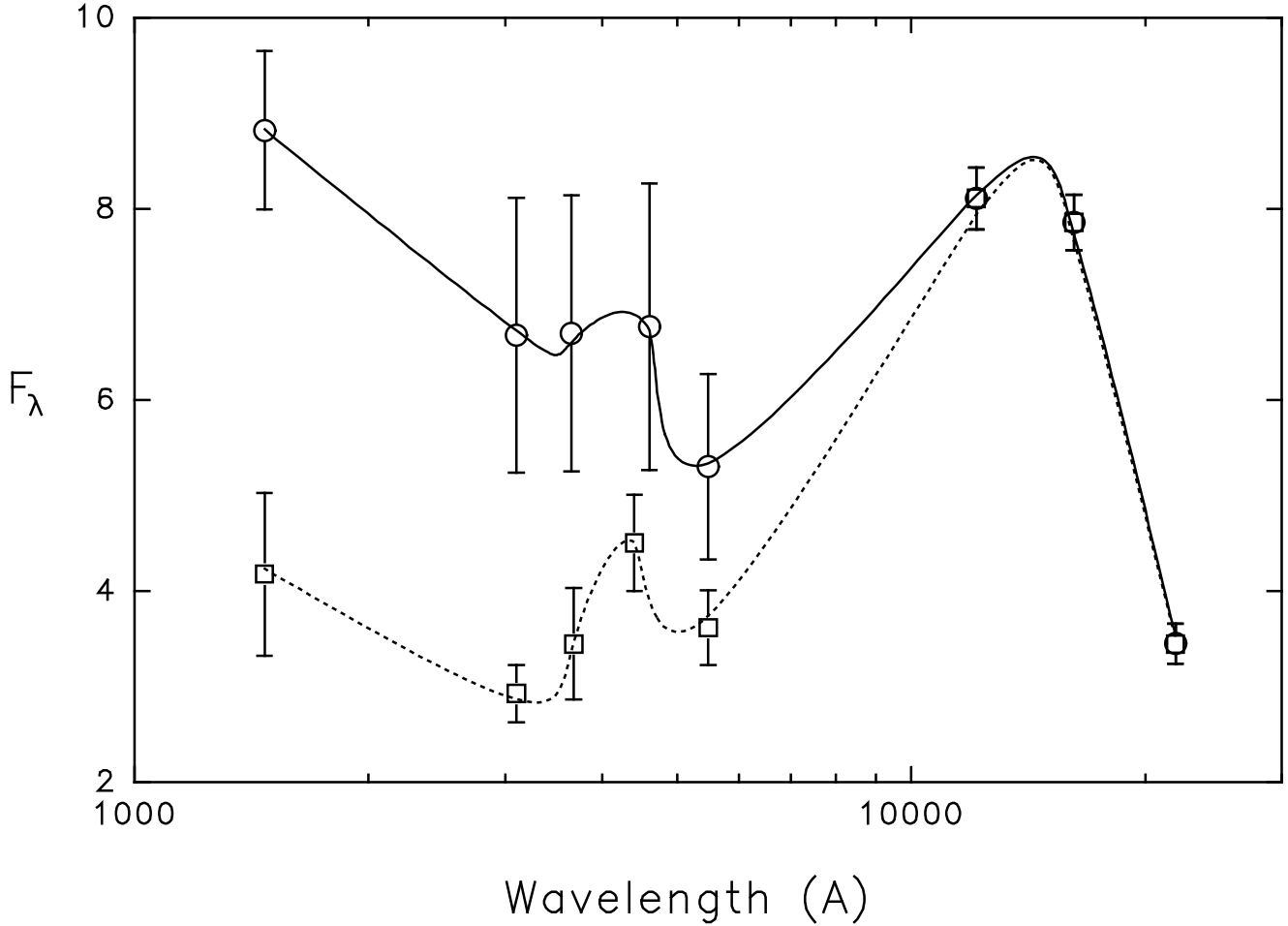}}
\caption{The continuum spectrum of MWC~560. Circles and
squares --- the data of observations in the active and passive states of the
object. Solid and dashed lines - the spectra of models: bb1\,(M5\,III) and
bb2\,(M5\,III). The wavelengths are in logarithmic scale. Intensity is in
${\rm 10^{-13}~erg~cm^{-2}~s^{-1}}$.} 
\end{figure*}

In our models the distance to the system is determined by the luminosity
of the giant star. However, this procedure is burdened with uncertainty
in absolute stellar magnitude of the giant. The distances we obtain are in
agreement with the results of Zhekov et al. (1996). The choice of the giant 
type (M4 or M5) has no essential effect on the parameters of the
hot component. The parameters of absorbing gas in the jet affect basically
the shape of the UV spectrum and the spectrum break near the Balmer
jump. This allows us to separate two effects: the jet gas absorption and
interstellar extinction. We have found the following parameters of gas
in the jet: T $\approx 7000\div 8000$~K, n $\approx 5\cdot 
10^{11}$~cm$^{-3}$, ${\rm N_H\approx 3\cdot 10^{23} \div 1\cdot 
10^{24}}$~cm$^{-2}$.  Approximately the same values for \nh\ have been 
obtained by Shore et al.  (1994) by modelling the absorption UV spectrum. 
From the found values of \nh\ and n we derive the jet length 
(of the absorbing region in the jet) to
be $3\div 6 \cdot 10^{11}$~cm in the active period and $1\div 2 \cdot
10^{12}$~cm in the passive. This is in good agreement with the constraint on
the jet length ${\rm R_j \la 2 \cdot 10^{12}}$~cm, found above from 
the time of variability of absorption lines in the passive period.

In the both models (ad and bb) the parameters of the hot source and jet
turn out to be quite close. We believe that the two models are not
alternative from the point of view of description of the object continuous
spectrum. A presence of accretion disk, at least its outer region, in 
MWC\,560 is quite possible. The central
region around the white dwarf may be screened from the observer by
an optically thick expanding wind, whose photosphere is the UV radiation 
source. The model bb is more self--consistent (see below). It can be concluded
from Table~2 that in both states the rate of mass loss is the same.
\nh\ in the active state is a few times less than in the passive one,
in the bb model this factor is 2. The gas velocity in the jet in the
passive state (the autumn of 1991) is about twice as low as in the
active. Since the mass loss rate is ${\rm \dot M_j \propto N_H\, V_j}$, both
the state do not differ in ${\rm \dot M_j}$ (unless the jet opening angle 
changes
in the different states). We have assumed above that all basic properties of
spectrum variations can be explained only by the variable jet velocity.
The assumption has been confirmed by the results of spectrum modelling.

When simulating the accretion disk spectrum we took into account only
the part of the disk that had a temperature higher than 10000~K. At lower
temperatures hydrogen is ionized only partially and a gas cooling in lines 
becomes
essential, which causes overestimation of the continuum level in the model.
In the accretion disk model the UV spectrum depends on \rin, \av(H) and
the jet parameters most of all. It appears that a fitted value \rin\ is close
to the radius of normal stars. Decrease in \rin\ causes increase in
bolometric luminosity and \nh, in particular for  \rin\ $= 5 \cdot
10^8$~cm (the radius of the white dwarf of mass ${\rm 1M_\odot}$) we have
\lbl\ $> 10^{39}$~erg/s and \nh\ $\approx 5\cdot 10^{24}$~cm$^{-2}$,
and the most luminosity is in hard UV range. The best fit of model spectrum
to the observed is achieved at \rin\ $\approx 4\div 8\cdot 10^{10}$~cm.
The accretion rate is determined unambiguously from the optical spectrum
and turns out ${\rm \sim 10^{-4}M_\odot}$/y; in the active state it is higher
than in the passive by about a factor of 1.5. The bolometric luminosity
of such a truncated accretion disk equals 1--3$\cdot 10^{37}$~erg/s.

The accretion disk model describes fairly the observed spectrum, but,
nevertheless, it is not suited to the system MWC\,560 since it requires
impossibly a high accretion rate and a large inner radius of the disk. The 
necessary accretion rate is ${\rm \sim 10^{-4}M_\odot}$/y. 
Besides that, it is unclear 
how the matter from the outer region of the disk is utilized. If this
matter outflows, then, in fact, this is the case of the bb-type model, in
which the required gas accretion rate is by two orders lower. The principal
problem for all the models --- the great UV luminosity of the source
and the relatively low accretion rate (${\rm \sim 10^{-6}M_\odot}$/y),
which may be provided by the red giant, can be naturally resolved in
the bb model. We will discuss therefore only this model of the hot
source --- the blackbody radiating sphere.

The two spectral types of the giant star considered M4 and M5 yield 
quite close parameters of
the source and jet. Nevertheless, it can be seen from Table~2 that the
value of interstellar extinction determined from the M5 giant spectrum is
systematically less than that found from the spectrum of the UV source.
The cool envelope of the giant may well contain a dust, therefore it would
be naturally expected that \av(H) $\le$ \av(G). This is indeed for
the model with M4 giant, so we consider it more preferable. The most 
likely parameters of MWC\,560 derived from the models
bb1 and bb2 $+$ M4 giant  are the next: the distance to the system 
D~$=1.4$~Kpc, the light extinction \av(H)~$\approx 1\magdot4$ and 
\av(G)~$\approx 1\magdot6$, the UV source temperature 
$\approx 20000$~K, its radius $\approx 4\cdot 10^{11}$~cm, its luminosity
 is $2\cdot 10^{37}$~erg/s in the active state and
$1\cdot 10^{37}$~erg/s in the passive state, the jet gas temperature
$\approx 7800$~K, its density on the line of sight is $3\cdot 
10^{23}$~cm$^{-2}$ in the active state and $6\cdot 10^{23}$~cm$^{-2}$ 
in the passive state.

The photosphere of an optically thick wind conforms to the blackbody
sphere model considered. This photosphere absorbs all direct photons 
from the white dwarf surface or from the inner regions, e.~g. from the 
magnetosphere. This may also account for the absence of a hard UV radiation
and high excitation absorption lines in the spectra of MWC\,560.
Such a wind may be arise 1) when the accretion luminosity 
exceeds the critical value, 2) as a result of thermonuclear burning
or bursts at the surface of the white dwarf (Pachynski and Zytkov, 1978),
and 3) as a result  of propeller mechanism action. The bolometric
luminosity of MWC\,560 is by an order lower than the critical one 
corresponding to electron scattering. If the hard radiation produced
 affects the unionized gas above
the accretion disk, the critical luminosity, needed for gas 
acceleration, may be markedly lower. Such a model requires
an axially-symmetric cocoon to be created above the internal parts of the
accretion disk, which, by itself, suggests the presence of other mechnisms
for destruction of the disk and gas acceleration. In continuous
termonuclear burning of matter accreted onto the surface of the degenerate
star, to ensure the observed luminosity ${\rm \sim 10^{-6\div -7} M_\odot}$/y
is needed, however the mechanism of gas ejection is absent. With
thermonuclear burst the luminosity in a burst may well be higher than
the Eddington luminosity necessary for outward acceleration of gas.
For the dramatic luminosity variations in the  bursts to be obscured
from the observer, the time intervals between the bursts must be much
shorter than the time of radiation emergence from under the
photosphere, i.e. ${\rm \ll R_{out}/V_j \sim 10^3}$ s. However the intervals
between the bursts with the accretion rates   being discussed
amount to dozens of years (Pachynski and Zytkov, 1978), therefore the
thermonuclear model of gas acceleration in MWC\,560 is not suitable.
We suggest that matter is accelerated by propeller mechanism.

In a number of papers devoted to investigation of MWC\,560  (e.g. Tomov
et al., 1994) it is assumed that in the passive state the white dwarf
accretes gas under propeller condition, while in the active state
(e.~g. in periastron the rate of accretion onto the magnetosphere
rises and, correspondingly, the magnetosphere size is reduced) the
star become the accretor. However it is not specified how matter is ejected
from the accretor. For instance, in the well--known polars, where gas is
accreted onto the surface of the white dwarf, no powerful, all the more
jet--like outflow of matter is observed. As we have seen, the gas accretion
rate in MWC\,560 is much lower than the critical, moreover, neither
luminosity nor gas accretion rate onto the magnetosphere changes
considerably in different states of this object. We have found above
that in different activity states of the object only the gas ejection
velocity changes essentially. Besides, in the active state the gas
velocity is strongly variable, accordingly the brightness variability
amplitude increases. What is more, in the active state the luminosity of
the UV source is two times higher. To achieve the observed luminosity of
MWC\,560 the required rate of accretion onto the surface of the white dwarf 
of a radius $5\cdot 10^8$~cm and of mass 1\,M$_\odot$ equals to 
${\rm \dot M_a= L\,R_{WD}/GM \approx 1.2\cdot 10^{-6}\ M_\odot}$/y in the
active state and ${\rm \approx 6\cdot 10^{-7}\ M_\odot}$/y in the
passive. When matter is propelled outwards from a magnetosphere, a part
of it may reach the white dwarf's surface, thus ensuring the observed
luminosity. We suppose that the white dwarf in MWC\,560 always
is under the propeller condition. We interprete the different activity levels
of the object as different propeller regimes: the active state of
MWC\,560 corresponds to the hard propeller, while the passive one 
to its soft regime. We will discuss this model in more detail below,
but first refine the parameters of the hot component, the jet and the system
MWC\,560.

\section{Parameters of the hot component and the jet}
It is highly probable that the UV radiation source around the white dwarf
of MWC\,560 is the photosphere of the powerful wind outflowing from the
internal regions. Let us discuss observational evidences for the existence
of this wind. The optical and UV spectrum is well described by blackbody
radiation from a spherical object with temperature of $\approx 20000$~K and
radius of $\approx 4\cdot 10^{11}$~cm (Table~2). Along with the detached 
high--velocity
variable absorption lines the low--velocity hydrogen absorption lines are
always present (irrespective of the state of the object (Tomov and Kolev, 
1997)), which make the line profile of the type P\,Cyg. The velocity of
outflow of gas in which these lines are formed is practically constant
and is $\approx -200$~km/s for the high members of Balmer series (Tomov
and Kolev, 1997). The depth of these lines is varied: in the passive
periods (for example, early 1991) the
high--velocity absorptions are getting closer to the low--velocity absorptions
and merge, the total absorption lines intensity increases.  The low--velocity
absorption lines may well be formed in the quasi--spherical wind being
discussed. The width of the narrow metallic emission lines as well
as the emission components of the high Balmer members is FWHM~$\approx
100$~km/s (Tomov and Kolev, 1997). These lines are probably emitted just
above the photosphere of the wind, the metallic lines are formed due to
fluorescence of UV radiation (Shore et al., 1994).

For the wind to be formed with the parameters --- luminosity of the
photosphere of $2\cdot 10^{37}$~erg/s, the temperature of 20000~K and
the size of $4\cdot 10^{11}$~cm, and the outflow velocity in the wind 
${\rm V_w = 100}$~km/s --- the required rate of the matter outflow is:

\begin{equation}
{\rm \dot{M}_W  \approx\frac{\textstyle 4\,\pi\, V_w\,
 R_{out}}{\textstyle k}\approx2.0\cdot10^{-6} M_\odot},
\end{equation}
where k~$\approx 0.4$~cm$^2$/g is the Rosseland absorption cross section
(Roger and Iglesias, 1992), which for the approximate parameters of
the photosphere, T~$= 20000$~K and $\rho = 4\cdot 10^{-12}$~g/cm$^3$ is,
in fact, equal to the Thompson cross section. The density adopted here
corresponds to the gas density at the distance \rou\ in the wind outflowing
at the rate ${\rm \dot M_w}$ and the velocity ${\rm V_w}$. Find the mass loss
rate in the jet and kinetic luminosity of the jet (the rate of kinetic
energy flux):

\begin{equation}
{\rm \dot{M}_j\approx\pi\,m_p\,\sin^2 \theta_j\, N_H\, R_{out}\, V_j
\approx 5\cdot10^{-7}\,M_{\odot}/y,}
\end{equation}
\begin{equation}
 {\rm L_k  = \dot{M}_j\, V_j^2/2 \approx 6\cdot10^{35}~erg/s.}
\end{equation}
Here we have used the MWC\,560 parameters in the passive state
(Table~2, bb2 model): the column density in the jet \nh\ $=6\cdot 
10^{23}$~cm$^{-2}$, the distance from the centre of acceleration to the 
jet base
is assumed equal to the photosphere radius \rou\ $=4\cdot 10^{11}$~cm, the
velocity of gas in the jet \vj\ $=2000$~km/s (Table~1), and the
semiopening of the jet $\theta = 20^\circ$. For the active state of the
object (model bb1) ${\rm \dot M_j}$ turns out to be the same as for the 
passive
state. As we have noted above, the mass loss rate in the jet as well as
the size and temperature of the photosphere of the hot component appear
to be constant (within the accuracy of the obtained model parameters) and
do not depend on the state of activity of the object. We consider this fact
to be of importance and it must be taken into account when selecting a model
of the object and discussing the gas acceleration mechanism. One of the 
most
important parameters, which characterizes the efficienty of the jet 
acceleration mechanism is the ratio of kinetic luminosity of the
jets \lk\ to bolometric luminosity of their source. In the case of
MWC\,560 the \lk/\lbl\ $\approx$ 0.05--0.1 depending on the activity
state of the object. If the gas outflow from the object was spherically
symmetric, the kinetic luminosity would then be close to the bolometric one,
which would be difficult to coordinate in the frames of the thermodynamics
in any model of gas acceleration. This attests independently the
the jet--like outflow of a gas in MWC\,560.

At the matter accretion onto magnitosphere under the proppeler regime 
a part of material may reach the star's surface. This is possible 
because of the instabilities in plasma -- magnetic field interaction,
because of a part of the matter falls in the
regions close to  the axis of rotation, etc. As we have seen
above, for the bolometric luminosity of MWC\,560 to be provided,
depending on the state of the object ${\rm \dot M_a \approx 0.6\div 
1.2\cdot 10^{-6}\ M_\odot}$/y must be accreted onto
the surface of the white dwarf .
Obviously, this is the upper limit of the rate of accretion onto the
surface since part of radiation (possibly, considerable) is produced
through heating and acceleration of gas by the propeller ${\rm L_{pr}=
L_k/\eta}$, where  $\eta$ is an efficiency of the propeller mechanism. We
can not estimate the value of $\eta$, most likely $\eta < 0.1$.  Even
in this case (see (3)) the propeller heating furnishes more than 30~\%
of the UV source luminosity. Thus, the total maximal rate of the gas flow
onto the MWC\,560 white dwarf makes ${\rm \dot M=\dot M_w+2\dot M_j+
\dot M_a \approx 3\div4 \cdot 10^{-6}\ M_\odot}$/y, or 
${\rm \approx 3 \cdot 10^{-6}\ M_\odot}$/y if the
luminosity of the objects provided by the propeller only. The main share
of the mass (about 50--70~\%) outflows at a small velocity (100--200~km/s),
forming the optically thick wind --- the source of UV radiation of MWC\,560,
$\la 15$\% of the total mass flow reaches the surface of the white dwarf and
approximately 20--30~\% is ejected as jets. We have obtained these values
from observations, they are rather rough, and some of them are 
model--dependent.

\section{The propeller in MWC\,560}
The presented gas stream pattern agrees fairly with the present--day
knowledge of the propeller mechanism operation at a gas accretion 
onto a rotating magnetic star (see the rewiev of the problem in 
Lipunov (1987)). This mechanism is very important for the study
of gas accretion onto fast--rotating neutron stars. This mechanism is also
effective in accretion of interstellar gas onto main sequence magnetic
stars (Fabrika and Bychkov, 1988). Now we have to discuss the MWC\,560 
binary system, i.~e. a possibility of accretion of the required
mass from the wind of the red giant, the parameters of the white
dwarf and it evolutionary status, and also possible mechanisms of
collimation of the jets and causes of the different states of the object
MWC\,560.

To satisfy the observed propeller luminosity and the rate of mass ejection,
the second star must lose not less than ${\rm \dot M \approx 3\cdot 10^{-6}\,
M_\odot}$/y. The average red giant mass loss rate in symbiotic variables is
${\rm \approx 10^{-6\div -8}\,M_\odot}$/y (M\"urset et al., 1991), 
but, probably,
even more in active states. The separation between the  components of MWC\,560
is ${\rm a=\left(\frac{\textstyle P_{orb}^2}{\textstyle
 4\,\pi^2}G(M_G+M_{WD})\right)^{1/3}} \approx 5.7\cdot10^{13}$~cm,
 where as previously, we take ${\rm M_{WD} = 1\,M_\odot}$,
 the orbital period of the system ${\rm P_{orb}=1930^d}$,
 and also the red giant mass ${\rm M_G}$, the mean value of which in
symbiotic stars of S--type is equal to 1\,M$_\odot$ 
(Whitelock and Munari, 1992). In long--period binaries the orbit is 
normally elliptical (Boffin et al., 1993).
The probable eccentricity value for the MWC\,560 orbital period may be
estimated at ${\rm e=0.15\,(\log\, P_{orb}-1)=0.34}$
 (Boffin et al., 1993). Then the minimal distance 
 between the components in MWC\,560 may be about
${\rm a_p=a(1-e)}\approx 3.7\cdot 10^{13}$~cm. 
Using this value and the expected mass ratio of the components
${\rm q=M_G/M_{WD}}=1$, derive the minimal critical radius  (Roche lobe)
${\rm
 R_{cr}=a_p(0.38+0.2\log\,q)\approx 1.4 \cdot 10^{13}~cm = 200\,R_\odot}$
 (Pachinski, 1971).
The typical radius of M4--5 giants is, however, about ${\rm 100 \,R_\odot}$
(Lang, 1992), i.~e. twice as small as ${\rm R_{cr}}$. The atmosphere of a
giant star is not stationary, it outflows irrespective of binarity,
the second star only intensifies the outflow. Therefore, and also taking
into account possible uncertainty in the parameters, we can conclude
that the giant star in MWC\,560 is capable of overflowing its
critical Roche lobe when the white dwarf is crossing periastron.

The gravitational radius of material capture by the white dwarf when
travelling on the orbit is ${\rm R_g=2\,G\,M_{WD}/(V_{WG}^2+V_{orb}^2)}$.
 The orbital velocity of the white dwarf,
as has been seen above, is $\approx 10$~km/s, the mean wind velocity of
the red giants is ${\rm V_{WG}}=10$~km/s (Nugis et al., 1989). At these values
of the parameters it is not difficult to find that ${\rm R_g \approx a}$. 
This means that independently whether the giant overfills its critical 
lobe, the white
dwarf may capture and accrete nearly all matter being last by the giant.
This inference can be backed up if one pays attention to the fact
that the orbital velocity is comparable in value with the velocity
of the wind outflow. The main portion of the gas lost can not leave the system,
it must be captured by the white dwarf.  From this, in particular,
the important conclusion follows that the rate of the gas accretion onto
the white dwarf may not be strongly varied with orbital phase. The
same conclusion has been drawn above, based on modelling the continuous
spectrum of MWC\,560. It is possible that in a system having such
parameters the accretion regime (disk or spherical) may alternate
depending on the orbital phase. For instance, at periastron the giant
may overfill its critical Roche lobe --- an accretion disk is formed,
in the opposite point of the orbit a white dwarf ''gathers'' gas of the
slow wind --- spherical accretion could be possible. To be precise, depending
on the phase of the orbit, changes the specific angular momentum of matter
captured by the white dwarf. So the pressure of gas falling onto the
magnetosphere is also changes, which may cause alternation of the propeller 
action conditions.

The propeller mechanism of a gas ejection by a star's rotating magnetic field
has been first suggested for neutron stars by Shvartsman (1970)
and has been developed  in a number of other works (Illarionov and
Sunyaev, 1975; Shakura, 1975; Halloway et al., 1978; Wang, 1979; Davies et
al., 1979, Davies and Pringle, 1981) (see Lipunov, 1987). Despite the vast
study of this problem no reliable propeller model has been created as yet.
The nonlinearity makes the problem  difficult to resolve. It is supposed
that zones of sectorial gas outflow are formed around the propeller,
or nonstationary gas ejection occurs when accretion is replaced by outflow.
In some propeller scenaries envelopes or disks may be formed round
the magnetosphere. It is evident that the propeller ejects gas and
carry away angular momentum of the star, however, to define concretely
the picture of accretion and outflow has been impossible so far.
For action of a propeller it is principally needed the excess magnetic
field rotation velocity at the boundary of the magnetosphere
(the magnetosphere size or Alven radius ${\rm R_A}$) over the velocity of
Keplerian rotation of matter at this radius, or, in other words,
${\rm R_A > R_c}$,
where ${\rm R_c}$ is the radius of corotation. Of vital importance
is the knowledge of the velocity of gas ejection. This is what
the propeller efficiency depends on, i.~e. the momentum loss
rate, the speed of evolution of stars etc. As a rule, two cases are
considered: 1) a hard propeller, when a gas is ejected by the magnetic
field at a velocity of its rotation (${\rm 2\pi R_A/P}$, where P is the star
rotation period) (Shakura, 1975; Holloway et al., 1978; Wang, 1979) and
2) a soft propeller, when the magnetic field heats a gas and it outflows
at a parabolic velocity (${\rm (2GM/R_A)^{1/2}}$) (Illarionov and
Sunyaev, 1975; Davies et al., 1979; Davies and Pringle, 1981). The magnetic
dipole is believed to approach a sphere through a dynamic pressing of
accreted gas, therefore the soft propeller scenario is considered as 
more preferable. However, 
Wang and Robertson (1985) have performed numerical simulations of the
propeller and shown that because of the Kelvin-Helmholtz instability
plasmoids may penetrate into a star magnetic field, i.~e. be carried along
by it and be ejected at a velocity close to the solid--body rotation 
velocity. Thus the
inclusion of possible instabilities in the propeller schema allows us to
expect that the hard version may also be realized.

We suppose that the white dwarf's magnetosphere in the system MWC\,560
undergoes transitions of the type ''hard-soft'' propeller. It is natural
to relate the active state (the high, several thousand km/s and variable
velocity of gas ejection) to the hard version, while the passive (the
outflow velocity is several hundred km/s) to the soft propeller version.
For the ''hard-soft'' propeller transitions no dramatic reconstruction of
the magnetosphere is required as is the case with the accretor-propeller
transitions (where even the hysteresis effects are expected (Lipunov, 1987)).
It is possible that with variations of the rate of gas accretion onto the
magnetosphere or with changing the condition of accretion (disk or spherical),
instabilities in the layer where magnetic field and gas interact, are
suppressed, which caused a change in the mode of gas ejection.

The size of the magnetosphere is determined by equality of the magnetic
pressure ${\rm B^2/8\pi}$
and the dynamic pressure of gas ${\rm \rho V^2}$, where V is a free-fall
velocity, and ${\rm \rho V = \dot{M} /
4\,\pi\,R^2}$. The dipole magnetic moment of the star is  ${\rm
\mu=B_0\,R_{WD}^3/2}$,
where ${\rm B_0}$ is the magnetic field strength at the pole. Assuming that
${\rm \dot{M} = 3\cdot10^{-6}\,M_{\odot}}$/y
 is accreted onto  the white dwarf's magnetosphere, derive the
magnetosphere size:
\begin{equation}
{\rm R_A=\left(\frac{\textstyle \mu^4}{\textstyle 8\,G\,M_{WD}\,\dot
 {M}^2}\right)^{1/7} \approx 1.2\cdot10^5 \,B_0^{4/7} \,\,cm}
\end{equation}
The corotation radius is determined by the rotation period P 
of the white dwarf:
\begin{equation}
{\rm  R_c=\left(\frac{\textstyle G\,M_{WD}\,P^2}{\textstyle
4\, \pi^2}\right)^{1/3} \approx 1.5\cdot 10^8\,P^{2/3}\,cm}
\end{equation}
From the propeller condition ${\rm R_A > R_c}$
we can find the relationship ${\rm B_0 > 8\cdot 10^8 (P/1000 \,s)^{7/6}}$~G.
The magnetosphere size is ${\rm R_A > 1.5 \cdot10^{10}}$~cm,
i.~e. the magnetosphere is completely
hidden from the observer beneath the photosphere of the outflowing wind
(${\rm R_{out} \approx 4\cdot10^{11}}$~cm).
For the propeller mode to be realized in MWC\,560, the white dwarf
must be strongly magnetized. The rotation period ${\rm P \sim 1000}$~s
corresponds to the
typical time of the observed brightness variability (10--20~min).
The rotating magnetic field is hidden well under the wind's photosphere,
and direct radiation does not reach the observer (it undergoes multiple
scattering and is thermalized). Nevertheless, ejection of gas in the active
state of the object may disturb the wind and carry the radiation outward.
So the object may be variable with a characteristic time of the order
of the rotating period of the white dwarf.

Let us estimate the magnetosphere size in other ways. The lowest 
observed velocity
of gas outflow in the passive state (Table~1) is about 600~km/s. Assuming
that in this state the propeller is in the soft mode, that is, the gas
outflow velocity ${\rm (2GM/R_A)^{1/2}}$,
find that ${\rm R_A \la 7\cdot10^{10}}$~cm. The inequality here indicates 
that, depending on the mass of the gaseous envelope lying on the
magnetosphere, the outflow may proceed at a velocity lower than the 
parabolic one. Take now that in the active state the matter 
ejection velocity in the jets
is about 5000~km/s.  If the assumption is made that in this state the
propeller operates in the hard mode, i.~e. the ejection velocity is close
to the solid--body velocity of magnetospheric rotation, find ${\rm R_A \approx
8\cdot10^{10}(P/1000\,s)}$~cm. The closeness
of the magnetosphere radius values obtained for the hard and soft modes
supports the conclusion that these modes do may exist, i.~e. 
the theoretical models of propeller.

Under the assumption that ${\rm R_A \la 7\cdot10^{10}}$~cm,
find the pole magnetic field strength of
the white dwarf, ${\rm B_0 \la 1\cdot 10^{10}}$~G.
From the condition ${\rm R_A > R_c}$ find the white dwarf maximal
possible rotation period,  ${\rm P < 2.8}$~hours.
The strong magnetic field of the
white dwarf in MWC\,560 is likely to be the necessary condition of the
gas ejection due to the propeller mechanism. The interval of magnetic field
pole strengths we have found, ${\rm 10^{9}\,G \la B_0 \la 10^{10}}$~G,
corresponds to the surface magnetic
fields of ${\rm 3\cdot10^{8}\,G \la B_S \la 3\cdot10^{9}}$~G.
Such magnetic white dwarfs are known (Schmidt et al., 
1990) among single objects. The frequency of white dwarfs
with such strong magnetic fields (Fabrika and Valyavin, 1997) is  about
5~\% among the magnetic white dwarfs. The magnetic field strength on white
dwarfs among polars (magnetic white dwarfs coupled with a dwarf of G--M
spectral type) makes $10^6 \div 10^8$~G.
For the white dwarf MWC\,560 evolution,
it is essential that this star is a wide pair with a giant. After completion
of the giant stage and formation of the second degenerate star, from
the observational viewpoint this magnetic white dwarf will not at all
differ (after the young white dwarf gets cooler) from a single star.

In the model discussed, the gas outflow in MWC\,560 will proceed until 
the giant phase is completed or the white dwarf slows down its 
rotation. The deceleration of magnetic white dwarf is the
most effective in the propeller state. The time--scale of the 
deceleration is determined by the forces moment that affect the
magnetosphere, which equals ${\rm 2 \pi \dot{M}_j R_A^2/P}$.
 The deceleration time is ${\rm t_d \approx M_{WD}R_{WD}^2/ \dot{M}_j R_A^2}$.
Assuming that in the most active state (as in the spring of 1990) the object
stays, on average, for about 10 \% of the total time and taking the above
limits to the magnetosphere radius, ${\rm 1.5\cdot10^{10}\,cm \la R_A \la
7\cdot10^{10}}$~cm, find the deceleration time
${\rm t_d \approx 10^3 \div 2\cdot10^4}$~years.
The time--scale of evolution of a giant at this
stage is about ${\rm M_{G}/ \dot {M} \sim 2\cdot10^5}$~years.
After formation of the white dwarf in MWC\,560,
it was at the propeller or ejecting stage (Lipunov, 1987). When the second
star of the system entered into the red giant phase, the rate of gas
accretion onto the magnetosphere of the white dwarf grew to
${\rm \dot {M} \sim 10^{-6} \,M_{\odot}/y}$,
and the white dwarf entered  the state being observed presently.
Several thousand years after deceleration of rotation of the white dwarf,
accretion onto the surface will be permitted, i.~e. the star will pass to the
accretor stage, at which secondary acceleration of rotation is possible.
The final state of the white dwarf rotation in the evolution scenario 
depends on
the state of the giant star. We can say that in any case from MWC\,560
a wide pair is formed, which consists of two white dwarfs. One of them
(magnetic) will rotate with a period from 10--20 minutes to a several
hours.

The propeller model in MWC\,560 allows both the nature of the
hot component of the system and its sporadic variability (flickering), and
the origin of the jets to be explained. From observations of MWC\,560
we know that jet--like outflow of matter occurs in the direction perpendicular
to the accretion disk (in any case it is along the angular momentum of 
the accreting matter). The opening of the jets is sufficiently large,
2\,\tetj \,$\ga 40^{\circ}$. So a more suitable
a term for the MWC\,560 jets would be the anisotropic wind or the sectorial 
structure of matter outflow described by Shakura (1975). One of the promising
models appropriate for the propeller is the collimation of jets with the aid
of magnetic field (see references in (Livio, 1996)). Blanford and Payne (1982)
have considered the picture of acceleration and collimation of gas by
magnetic field above the accretion disk. This picture may be suited to the
case of collimation of gas ejected by the propeller. Gas is entrapped
by the magnetosphere magnetic field, in so doing the field is distorted,
the open magnetic lines form a toroidal field component. Above 
the magnetosphere
the toroidal field gets predominant and may accelerate and collimate gas
along the axis of rotation. It does not seem possible to give a more specific
description of the picture of acceleration and collimation of gas in MWC\,560.
The bulk of observational evidence permits one to hope that the object
will stimulate the further development of the propeller model.

\section{Conclusions}
Spectroscopy results obtained at the 6-m telescope for the unusual symbiotic
star MWC\,560 in the period from the autumn of 1990 to the spring of 1993 are
presented. The high-velocity absorption lines of H\,I, D$_1$\,D$_2$~Na\,I,
Fe\,II(42) and He\,I, which radial velocity varies with a
time--scale of several months and is equal to $-500 \div 
-2500$~km/s are permanently present in the spectrum. 
The shape of the high--velocity
absorption profiles in the spectrum of MWC\,560 suggests that these are
formed in a jet directed towards the observer. These absorption lines are
variable at times shorter than one day with an amplitude of about 10~\%. A
fast variability for a time of $\la 3$ hours has been detected. Evidence
has been found for the existence of an opposite jet directed away from the
observer: in the red wing of the \hal\ line there is extra and 
variable radiation,
whose radial velocity corresponds to the velocity of matter ejection in the
jets measured from the absorption lines.

From the constancy of emission line radial velocities the inclination angle
i\,$\approx 10^\circ$ has been estimated. The continuous spectrum has been
modeled  from the UV to IR range in two different, active and passive,
states of the object. The observed spectrum is formed by a star of
spectral class M4--5\,III, a hot source and an absorbing screen (the jet).
The most likely spectral class of the giant star is M4. The interstellar
extinction to MWC\,560 is \av\ $\approx 1\magdot4$, the extinction to the
giant star may be by 0\magdot2 greater. Parameters of the MWC\,560 jets
and physical conditions of gas in them have been estimated. The opening
of the jet is ${\rm 2\theta_j \ga 40^\circ}$, its length at the place 
where the
absorbent is located is $\approx 1\cdot 10^{12}$~cm. The temperature and the
density of gas in the jet are ${\rm T_j \approx 8000}$~K and n~$\approx 
5 \cdot 10^{11}$~cm$^{-3}$. The density on the line of sight is varied, 
${\rm N_H \approx
3\div6 \cdot 10^{23}}$~cm$^{-2}$. In different activity states of the object
only the velocity of gas ejection (and accordingly the value of ${\rm N_H}$) 
 changes
at an approximately constant mass loss rate in the jet, ${\rm
\dot{M}_j\approx5\cdot10^{-7}\, M_{\odot}/y}$. The luminosity
of the object ${\rm L_{bol} \approx 2\cdot 10^{37}}$~erg/s
is basically determined by  the UV source with the temperature
T~$\approx 20000$~K, whose radius is ${\rm R \approx 4\cdot10^{11}}$~cm.
The hot source is, apparently,
the photosphere of strong wind (${\rm
\dot{M}_j\approx2\cdot10^{-6}\, M_{\odot}/y}$)
outflowing from the region around the
white dwarf at a velocity of 100~--~200~km/s.

We suppose that the wind and the jets are accelerated in the magnetic field
of the white dwarf which accretes gas from the wind of the giant under the
propeller regime. The gas is supplied through the accretion disk and ejected
by the rotating magnetic field. A sectorial structure of the gas outflow 
(the jet) is formed. The different activity states of the object may
correspond to different modes: a hard propeller with a gas ejection velocity
of several thousand km/s and soft propeller with an ejection velocity of
hundreds of km/s. The total rate of gas accretion onto the propeller is
${\rm \dot{M}= 3 \div 4\cdot10^{-6}\, M_{\odot}/y}$. The main part 
of it (about 50--70~\%) outflows as a slow wind.
Not more than 15~\% of gas accreted onto the magnetosphere may reach the
surface of the white dwarf. This share depends on which proportion of the
object's luminosity (may be the whole luminosity) can be provided by the
propeller. About 20--30~\% of the initial captured mass is ejected as jets.

For the effective operation of the propeller, the white dwarf in MWC\,560
must be strongly magnetized. The strength of its surface magnetic field may
be ${\rm 3\cdot10^{8}\,G \la B_S \la 3\cdot10^{9}}$~G,
the period of its rotation may lie within the interval from
10--20 minutes to 3 hours. The lower limit of the period is close to the
scale of photometric flickering of the object, therefore the period of
its rotation is likely to be about 10--20 minutes. The magnetic field strength
on the magnetosphere may amount to $\sim 10^4$~G.
The magnetosphere is completely
hidden from the observer under the photosphere of the slow outflowing wind.
Direct observations of radiation from the magnetosphere or from the surface
layers are impossible since it is scattered much once before it appeares
on the photosphere. It is possible that in active periods, when gas is
ejected in portions at a velocity of several thousand km/s, channels may be
formed in the gas envelope through which the observer can recognize bursts of
hard or polarized radiation. Further spectral and photometric observations
of MWC\,560 are of great importance.

\begin{acknowledgements}
\noindent The authors are grateful to D.~Kolev, E.\,L.~Chentsov and S.~Popov
for helpful discussions, A.\,I.~Shapovalova and A.\,N.~Below for assistance
in observations, S.~Sergeyev and G.~Galazutdinov for programmes for
reduction of spectra. A.A.P. thanks the Bulgarian National Astronomical
Observatory ''Rozhen'' for welcome during the work over the paper. The work
has been supported by the RFBR through grant 96-02-16396. Thanks of T.T.
are due to the Bulgarian Foundation of Scientific Researchs for support
by grant F-35/1991.
\end{acknowledgements}

\end{document}